*electronics* MDPI

*Article*

# Inverted Microstrip Gap Waveguide Coplanar EBG Filter for Antenna Applications

Luis Inclán-Sánchez

Department of Signal Theory and Communications, University Carlos III of Madrid, 28911 Leganés, Spain; linclan@ing.uc3m.es

**Abstract:** The possibility of making compact stopband filters using coplanar-coupled EBG resonators in inverted microstrip gap waveguide technology is studied in this work. To do this, the filtering characteristics of different configurations of mushroom-type elements are shown in which the short-circuit element is placed on the edge of the resonator's patch. The behavior of the structure as well as its main advantages such as: low losses, self-packaging, low level of complexity, flexibility and easy design are illustrated in the paper. To evaluate the possibility of integrating these structures in gap waveguide planar antennas feeding networks, a 5-cell EBG filter was designed and built at the X band. The proposed filter reached a maximum rejection level of −35.4 dB, had a stopband centered at 9 GHz and a relative fractional bandwidth below −20 dB of 10.6%. The new compact filter presented a flat passband in which it was well matched and had low insertion losses that, including the connectors, were close to 1.5 dB in most of the band. These results are enough to improve low-complexity future antenna designs with filter functionalities in this technology.

**Keywords:** gap waveguide technology; inverted microstrip gap; stopband filter; metamaterial; electromagnetic bandgap structure; artificial magnetic surface; antenna feeding network; antenna array; X band

## 1. Introduction

Today in our society there is an intensive use of wireless communications systems that needs to be managed in many aspects. The first consequences are a growing saturation of radio spectrum resources and the need to increase communications data rates, causing the extensive use of higher frequency bands, i.e., mmWave bands. On the other hand, a balanced distribution of both energy and interference levels is necessary to enable wireless connections between devices in this radio ecosystem, including new cognitive spectrum sharing methods with high requirements on latency and reliability [1].

A key aspect in the devices that will be required in the future will be the integration of filtering capabilities that allow operation in multiple bands. This is not a new requirement; in the past very high isolation levels have been achieved with the integration of filters in the antenna's own feeding network [2]. However, current radiant systems increasingly require the ability to operate in more distant frequency bands, use a smaller space for the antenna and address greater complexity using several technologies. For example, in [3] the possibility of configuring a MIMO array in a Smartphone that works at 3.5 GHz, 28 GHz and 38 GHz frequencies was recently demonstrated by integrating an isolation filter.

In this context, in order to reduce losses in high frequency bands, various improved technologies have emerged in the last two decades such as substrate integrated waveguide [4] and gap waveguide [5]. We are going to focus on the gap waveguide technology that in some applications can offer important advantages to integrate advanced low-loss antenna array systems operating at frequencies above 10 GHz [5–8]. This technology avoids the use of substrate or electromagnetic propagation through it, which eliminates the losses associated with them. It is also mechanically robust, offers an inherent condition of packaging, the guided structure does not require electrical contacts between the layers and its manufacture is becoming easier and consequently cheaper. However, it also has drawbacks, for example it occupies a considerable amount of volume and it can be heavy, even though recent studies have shown that there are 3D printing manufacturing solutions that can reduce the weight [9,10]. On the other hand, sometimes the integration of feeding networks is not easy since their size is not compatible with the separation required by the antennas in the array. For instance, it is sometimes necessary to carry out complex transitions between different versions of the gap waveguide technology [11], while in others the radiating device needs multiple layers to enable adequate power distribution [12].

Gap waveguide technology has been used successfully to design passive elements, and much attention has been focused on filter design in different versions of the technology such as groove and ridge [13–15]. New combinations of these versions have recently been proposed that integrate the microstrip line with the ridge concept of the gap waveguide [16–18]. The number of works that apply the inverted microstrip version for filter design has been more limited; however, we can review interesting examples. In [19], an end-coupled bandpass filter at 60 GHz was implemented, demonstrating the low losses that this version of the technology can offer. Other examples of filters in this technology can be seen in [20], which offer a good performance at a frequency of 28 GHz. Some of these filters are also being integrated for the implementation of diplexers, both in versions without substrate [21] and in inverted microstrip type [22].

In this work, the electromagnetic bandgap periodic structures concept is used to produce frequency stopband. These EBG structures have been applied in the past in many applications, such as for designing stopband and lowpass filters [23–26]. Some of these proposals, in their 1D and 2D versions, have been used in improved antenna designs, allowing high isolation [2,23], or reducing surface wave excitation [27–29].

Applying this methodology in a simple way, our purpose was to design compact filtering structures that are easy to integrate into new versions of planar antennas in gap waveguide technology. These antenna array systems with new filtering capabilities would enable multiband operation or interference mitigation. By means of a novel configuration of coplanar compact resonators that load the microstrip line, the possibility of making filters of minimal complexity will be demonstrated. The main advantage of these filters is that their design is independent of the initial feeding network and their manufacture essentially affects the printed circuit board of the line. This may allow reusing, optimizing and cheapening the manufacture of the artificial magnetic surfaces used in the gap waveguide. Although it has been sufficiently demonstrated that this technology offers low losses in these frequency ranges (X, Ku, Ka bands and beyond) [19,20,30], losses associated with the propagation of the signal in this filter at higher frequencies is another relevant aspect that will be addressed in the study. Finally, the design flexibility and easy frequency reconfigurability offered by this structure will be illustrated.

## 2. Filter Characteristics and Design

An approach to the design of microstrip filters is their treatment as EBG structures, in this case the elements periodically load the transmission line to obtain the stopband characteristic [23,24,26]. In the case of microstrip technology, a type of resonator that has been widely used to design filters is the mushroom element proposed in 1999 [27]. Some of these designs place the resonators underneath the transmission line in a multilayer configuration [2,31,32]. This topology would not be useful in the inverted case since another additional layer of substrate would be necessary through which the electric field would propagate. In this way, the losses due to the substrate would be increased and the required air signal propagation, which is an essential advantage of this technology, would be lost. This is the starting point of this work to investigate the option proposed in [33–36], coupling the resonators in a coplanar configuration to the inverted printed line on the same circuit board. In this case the mushroom elements were short-circuited to the top plate of the parallel plate waveguide in an inverted position as shown in Figure 1.

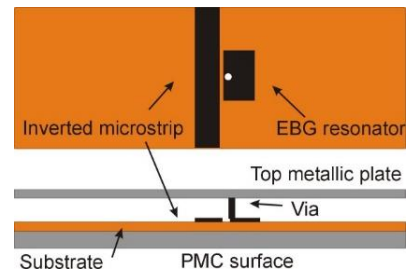

**Figure 1.** Proposed configuration for the coplanar loaded resonator in inverted microstrip gap waveguide technology.

Throughout the development of the work, the CST Studio Suite was used as a full-wave electromagnetic simulator, both for the simulation of S parameters with truncated 3D structures and for obtaining the dispersion diagrams of the ideal periodic structures.

*2.1. Artificial Magnetic Surface*

As the artificial magnetic surface, we used a square arrangement of pins, called a bed of nails in the literature [5,19]. Its operation is based on the creation, through the metallic pins, of a high impedance boundary condition. In our case, simulations were carried out until a response in the band of interest was obtained. Figure 2 shows the parameters of the considered cell and the dispersion diagram obtained. It can be seen how this configuration prevents the propagation of parallel-plate modes in a frequency range between 7.4 GHz and 14.8 GHz, which is what we call the stopband. In this frequency range, the textured surface behaves as a perfect magnetic conductor, and if the distance between the parallel plates of the waveguide is less than $\lambda/4$, it prevents the propagation of electromagnetic energy [5,6,19]. It is evident that in this case the electric field cannot fulfill the boundary condition between two perfect conductor surfaces, one electric and the other magnetic. This condition is inherently broken, in this specific case by means of an inverted microstrip transmission line as shown in Figure 1, which enabled the propagation of the field. In this configuration, a quasi-TEM mode was excited and confined in the air gap between the upper plate and the printed line above artificial magnetic surface.

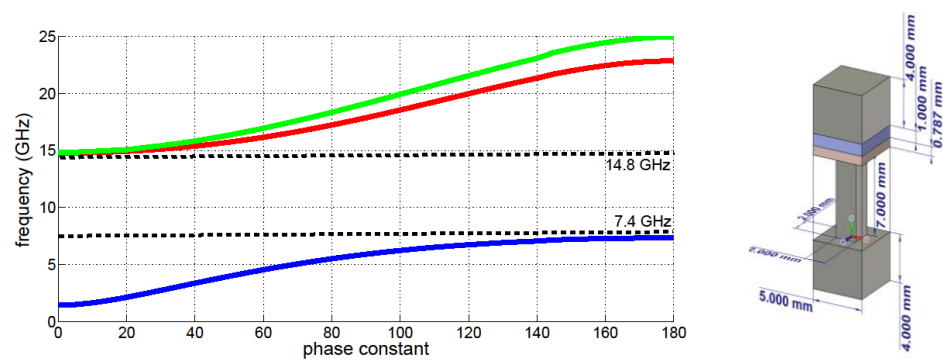

**Figure 2.** Periodic cell for the bed of nails and calculated dispersion diagram for the array of pins.

*2.2. Resonator Filtering Characteristics and Periodic EBG*

The resonator proposed here loads the line in a coplanar configuration, producing a narrow frequency notch in the passband. In the case of a single resonator, the filtering frequency depends on its geometric elements. First, the width and length of the printed patch and its gap to the inverted microstrip line establish the resonance condition and determine the rejection frequency. Secondly, the frequency is affected by the radius of the short-circuit element (metallic via) that electrically connects it to the top plate of the Gap waveguide, and it is also influenced by the position of the metallic via with respect to the patch. All of them are important, and as we will see throughout the paper, they also allow the element's working frequency to be adjusted flexibly. Given that the objective is to have filtering structures that can be easily integrated into feeding antenna networks, the resonator size is an essential requirement, which is why the goal is to obtain compact filters. It is well known that the size of this type of resonator can be reduced considering the lateral position of the metallic via, that is, placing it right on the edge of the patch [37]. On the other hand, the resonator can be miniaturized by increasing the capacitive coupling between the patch and the line. This means that we can achieve a smaller size resonator considering patches with rectangular geometry with the long side coupled to the microstrip line. This will be the case developed throughout the work. Finally, an important characteristic is the inductance created by the metallic via (or metallic post), the patch could be made smaller as its radius decreases. These options to make the proposed EBG cell more compact are outlined in Figure 3.

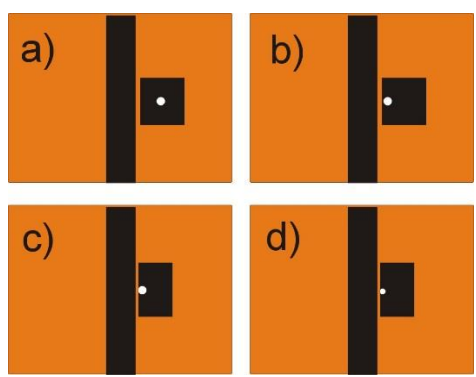

**Figure 3.** Strategy for EBG resonator miniaturization, steps to compact the element: (**a**) square patch with centered metallic via, (**b**) square patch with lateral metallic via, (**c**) rectangular patch with longer capacitive coupling and (**d**) metallic cylindrical shorting element (metallic via) with lower radius.

For the development of the numerical study, a 100 mm × 50 mm truncated bed of nails was chosen (see Figure 4a). The textured surface was formed by the 2 mm × 2 mm square pins, shown previously in Figure 2, with a period of *p* = 5 mm and a height of h = 7 mm. The substrate, in this case a Rogers 4003, was placed on the pins mechanically supported by them and the copper microstrip line was placed on top of it. The thickness of the substrate was hs = 0.79 mm, its permittivity was ε = 3.38 and its losses at 10 GHz were given by tg δ = 0.0027. Between the microstrip line and the metal top parallel plate there was a 1 mm thick layer of air. The line allows the propagation of the electromagnetic field between its surface and the parallel plate cover, so that the transmission of energy occurs essentially in air, which reduces losses in this gap waveguide technology. The structure remained fixed for all simulations and was the one manufactured in the results part of the paper. Of course, the thickness of the air layer affects the response of the proposed structure. As it was intended to analyze the parameters that directly affect the geometry of the resonator, throughout the study we considered the same thickness hair = 1 mm for this layer of air. It should be noted that it would be possible to modify this variable always respecting the condition of maximum distance between the upper plate and the textured surface that ensures the operation of the gap waveguide. With a greater thickness of the air layer, which is what fundamentally propagates the electromagnetic energy, it would be possible to transmit higher levels of power.

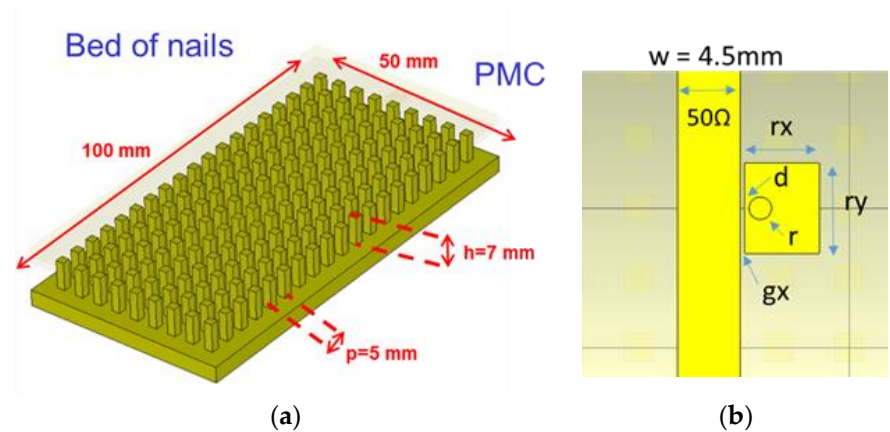

(a)          (b)

**Figure 4.** Truncated bed of nails for analysis and main parameters for the EBG resonator: (**a**) array of pins (**b**) EBG resonator in coplanar configuration and geometrical parameters.

We will start the analysis showing the simulation in this structure of a line-coupled resonator as shown in Figure 4b. Once we specify the dependence on the response with its main parameters, we will extend the study to filters formed by three and five EBG elements. Figure 5 shows the simulations for the case of a squared patch with the via in central and lateral position, and it can clearly be observed that for the same patch size, the resonance for the lateral case occurred at lower frequencies, which allowed the element to be compacted. Another possibility is to increase the length of the capacitive coupling of the patch with the line; in this case it can be seen how increasing the capacitance (the coupling length between the microstrip and the patch) allowed a resonance at a lower frequency. Finally, a critical parameter, as we will analyze later, is the radius of the short-circuit element. In the simulations this element (metallic via) was modeled as a metallic cylinder of length always constant and equal to the air layer hair = 1 mm. It can be seen in Figure 5 the considerable effect of the radius (r) on the resonator response. In the case of the rectangular resonator (see Figure 5) a variation from r = 0.8 mm to r = 0.6 mm reduced the resonance frequency from 10.42 GHz to 9.64 GHz. This variable, with its practical limitations, made the filtering response of these EBG structures more flexible, as we will see in detail in the results section of the paper.

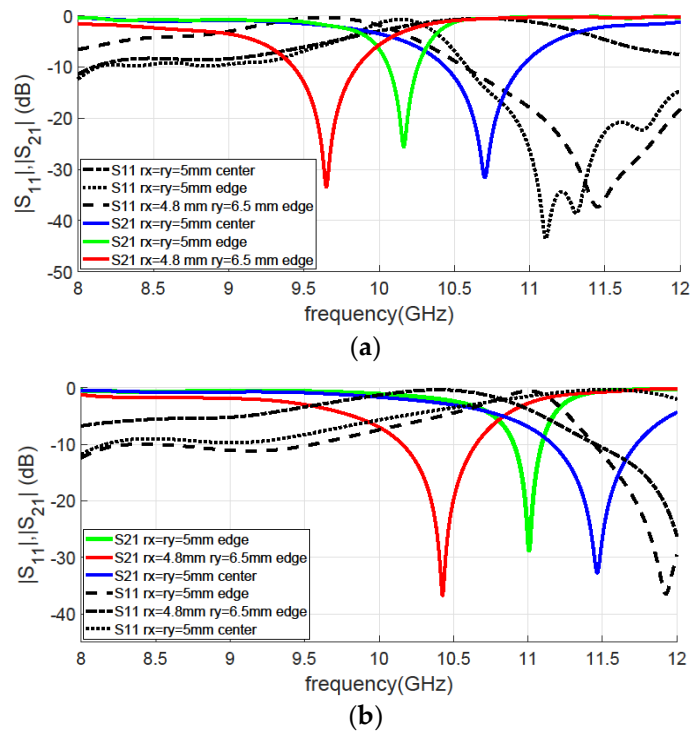

**Figure 5.** Simulated resonator response for different topologies (gx = 0.2 mm); square patch with central metallic via, square patch with lateral metallic via and rectangular patch with lateral metallic via with: (**a**) r = 0.6 mm and (**b**) r = 0.8 mm.

Figure 5 clearly shows the narrow frequency notch in the passband produced by the resonators depending on their size and the position of the shortcircuit element. To illustrate this effect, the simulated electric field distribution for the resonator of Figure 4b, with rx = 4.8 mm, ry = 6.5 mm, gx = 0.2 mm and r = 0.8 mm, at two frequencies is included in Figure 6. At the resonance frequency f = 10.45 GHz, the effect of the cell is to prevent the propagation of energy, while outside this stopband (at f = 11 GHz in Figure 6b) the field can be transmitted locally guided by the inverted microstrip provided that the pin surfaces are working in their bandgap, that is, they behave as an equivalent magnetic surface. Otherwise, the field in the parallel plate would be excited.

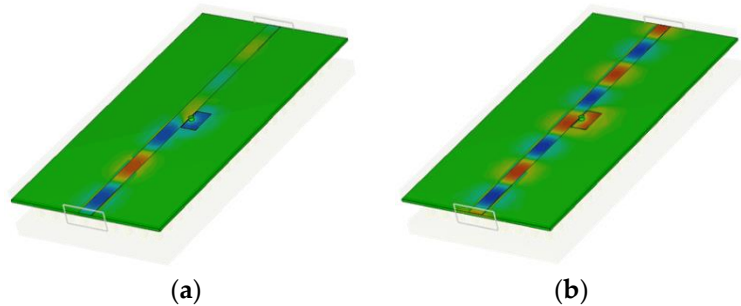

(**a**) (**b**)

**Figure 6.** Simulated Ez field component for the 1 cell resonator (rx = 4.8 mm, ry = 6.5 mm gx = 0.2 mm and r = 0.8 mm) response at frequency (**a**) f = 10.45 GHz (**b**) f = 11 GHz.

The parameters that most affect the size of the resonator are the dimensions of the patch. We have previously discussed our choice of rectangular geometry patches that allow for greater capacitive coupling. In order to clearly appreciate the effect that the length (ry) has on the resonance frequency, the simulations for different sizes of the length of the element are included in Figure 7, leaving its width fixed at a value of rx = 5 mm. It can be seen how as the length increases, the resonance frequency decreases due to an increase in coupling. This would mean a higher capacity in the circuit model proposed in the literature for this type of resonator [31–33]. A parametric study to systematically analyze the variation of the resonance frequency of the element is shown in Figure 8.

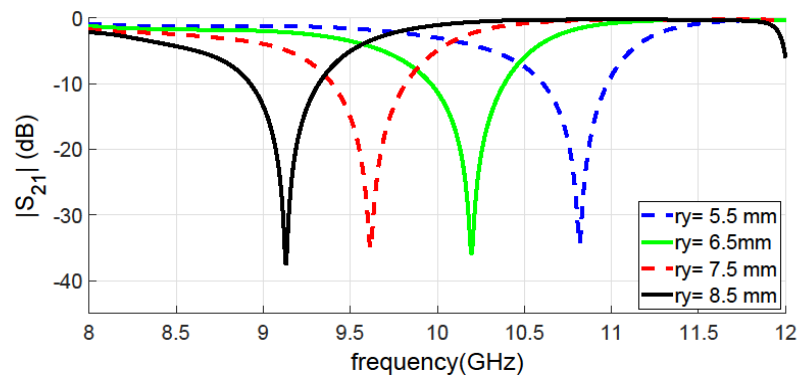

**Figure 7.** Simulated resonator, with via in the lateral position, transmission response for different length (ry) in mm (rx = 5 mm, gx = 0.2, r = 0.7 mm, d = 0.3 mm).

Figure 8a summarizes the previous effect of size and describes the significant variation that the radius of the short-circuit element had on the filtering frequency of the resonators. This variation was appreciably larger than originally expected and has not received much attention in previous published studies of this resonator in the same configuration. Clearly, as the cylindrical metal post reduced its radius, the resonance frequency decreased. This allowed the size of the elements to reduce in a simple way in case this was a necessary design requirement, and to easily adjust the response by exchanging the short-circuit elements. This parameter, in addition to adjusting the working frequency of the filters, offers a high degree of flexibility in order to design simple filters that can be integrated into the feeding network of an antenna array.

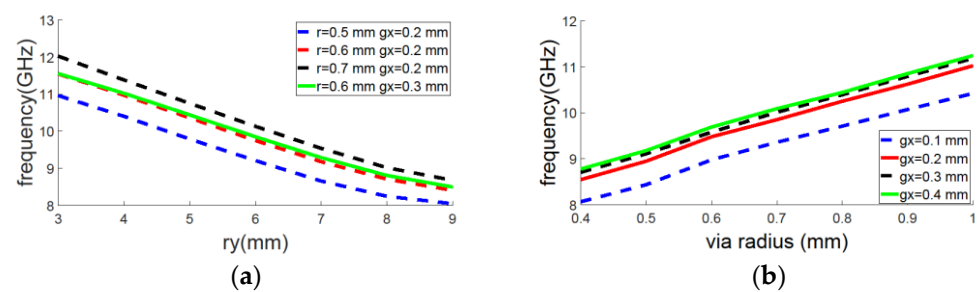

**Figure 8.** Resonant frequency for the EBG resonator loading the line as a function of: (**a**) resonator length (ry) (for differents radius (r) and for two gap values with rx = 5 mm) (**b**) via radius (r) (for differents gaps with the line (gx) and rx = 5 mm).

In this way, Figure 8b shows the wide range of variation of the operation frequency that can be obtained in the X band with short-circuit elements with a radius between r = 0.5 mm and r = 1 mm. Another aspect to take into account is the distance between the resonator and the microstrip line. In this frequency range, between 8 GHz and 12 GHz, a small gap between the line and the resonator is necessary to ensure coupling. For example, in our case for frequencies around 10 GHz we considered gaps that varied between 0.1 and 1 mm. Figure 8b shows that for very small gaps the coupling was large, the capacity increased, and the resonance frequency decreased. For gaps between 0.1 mm and 0.3 mm, the behavior was adequate and allowed sufficient coupling, and for higher values the coupling decreased significantly. To avoid manufacturing tolerance problems in our laboratory, a small gap of gx = 0.2 mm was chosen for most of the work.

## 2.3. Truncated EBG Version and Filter Design

In this section we made a version of the filter by cascading several resonators loading the microstrip line [2,18,24,36]. EBG structures offer higher levels of rejection as the number of cells or resonators increases. That is, as the order of the equivalent filter grows. The increase in the number of elements also affects the ripple in the pass band, although in our case the reduced number of EBG cells limited this ripple in the response [24–26]. A number of three elements was chosen to study the basic response of the filter to the most relevant parameters previously identified. We can see schematically in Figure 9 the layers that made up the gap waveguide used as a model.

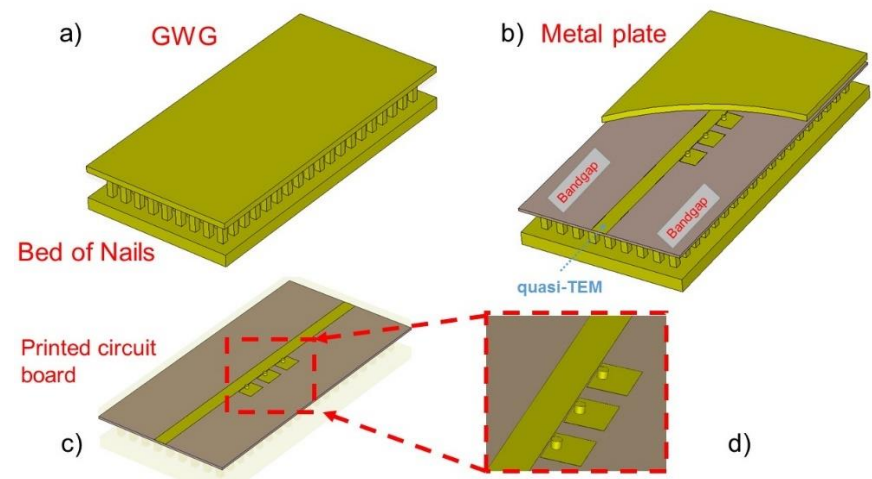

**Figure 9.** Proposed inverted microstrip gap waveguide configuration for the filter with the truncated EBG structure (3-cells): a) View of the parallel plate gap waveguide b) Multilayer configuration for the proposed inverted microstrip c) Printed circuit board with the filter d) Detail of the EBG resonators

Figure 10a shows a front view of the parallel plate waveguide with the detail of the air layer thickness. Figure 10b indicates the most important parameters for the characterization of the response of the EBG structure in the performed parametric analysis. Note that since the filter now has several elements, a new parameter (gy) indicates the distance between resonators. It is clear that the period of the periodic structure in this case is given by the sum of the length of the resonator patch (ry) and the distance between cells of the periodic structures (gy). This parameter determines the frequency response of the filter, although its effect, in the numerical analysis carried out, is not the most relevant. It should be noted that if the resonators are very close, the capacitive coupling between them increases, which makes it more difficult to adjust the filter response and complicates the design. For this reason, a somewhat greater distance was chosen, reducing the coupling between resonators, which allowed a sufficient stopband and at the same time a simpler interpretation of the simulations.

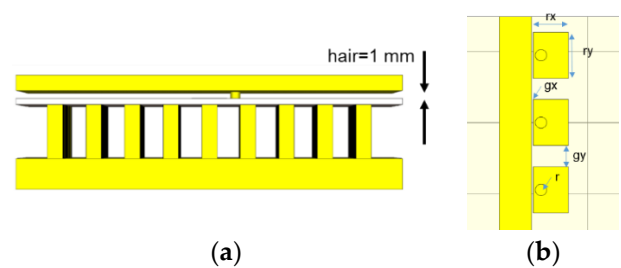

(**a**)　　　　　　　(**b**)

**Figure 10.** Details of the models used for the simulations: (**a**) Front view of the parallel plate with the thickness of the air layer (**b**) Main design parameters of the EBG filter.

The first aspect of the filter's frequency response previously analyzed is shown in Figure 11a. As the resonator patch length increased, its operating frequency reduced. Given a resonator patch width rx = 5 mm and varying its length from ry = 4.5 mm to 6.5 mm, the filter response shifted from f = 11 GHz to f = 9.5 GHz. Of course the radius of the short-circuit element produced a significant effect that we have already discussed and that we will analyze again for the case of three resonators. On the other hand, a parameter that we introduced to ensure mechanical tolerances (the displacement of the via from the edge of the patch (d), see Figure 4b), had a remarkable effect on the response of the filter. The size of the patch was somewhat larger, but as we can see in Figure 11b small variations of this distance (d) can produce, in addition to stopband shifts towards higher frequencies, a widening of the filtering bandwidth that may be interesting in some cases.

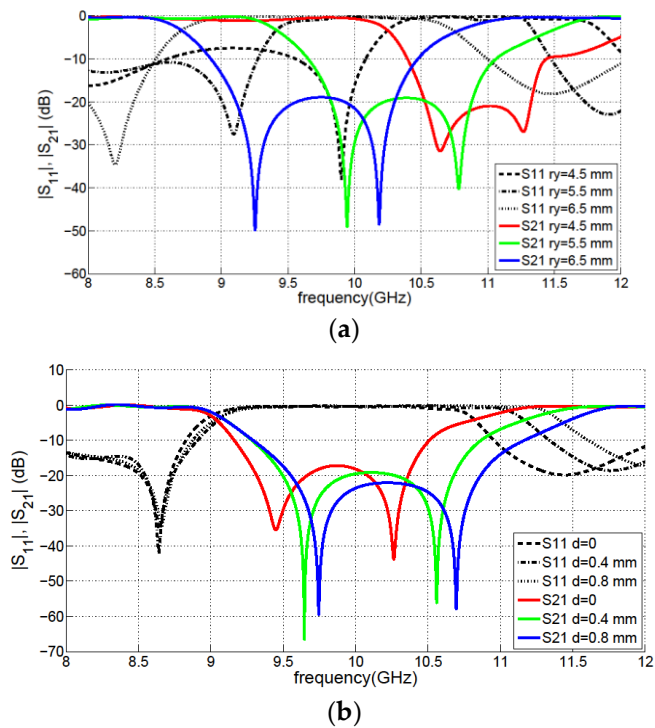

**Figure 11.** Simulated S21 response for the 3-cell filter (rx = 5mm, r = 0.75 mm, gx = 0.2 mm and gy = 2.5 mm) varying: (**a**) coupling length (ry) (**b**) via position distance (d) from the edge of the resonator patch.

A parametric study of the effect of the main parameters of the 3-cell filter on its frequency response was carried out. The synthesis of the main results is shown in Figure 12. We focus in this analysis on the effect that the radius of the short-circuit elements and the gap between the resonators and the microstrip line have on the response.

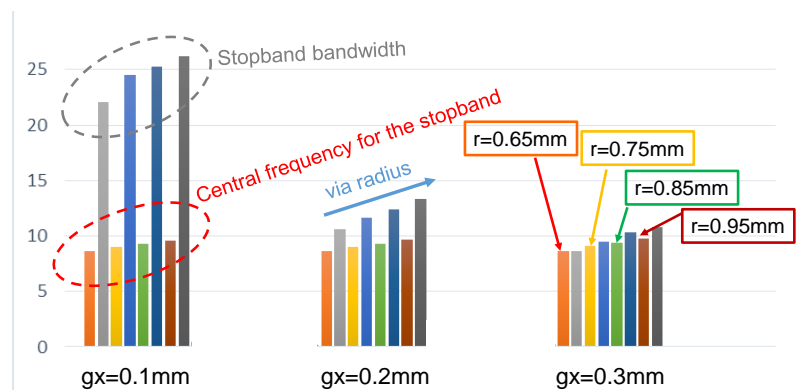

**Figure 12.** Stopband central frequency and its fractional bandwidth (assuming a rejection level greater than 10 dB) as a function of: the via radius (r) and the gap between the patch and microstrip line (gx). Other main parameters (rx = 4.9 mm, ry = 6.5 mm, gy = 2 mm, and d = 0.3 mm).

The image shown in Figure 12 represents for different values of the radius of the short-circuit element (r) and for three values of the gap between the resonators and the transmission line (gx); the stopband central frequency obtained by the filter and its fractional bandwidth considering a rejection level greater than 10 dB.

The main useful conclusions for the design of this type of filters that we can obtain from the analysis are: The first, already known, is that as the radius of the short-circuit elements decreased, the working frequency of the filter decreased. Secondly, it was also observed how, in a systematic way, a reduction in the radius of the via reduced the achievable bandwidth for the stopband. Finally, the small variations simulated for the gap between the line and the patch had no appreciable effect on the operating center frequency. However, they produced a considerable increase in the stopband bandwidth as the gap decreased, as shown in Figure 12.

## 3. Results and Discussion

Once we analyzed the behavior of the filter based on its parameters, this section shows the simulation and measurement results of a demo filter that evaluated the use of this new configuration for antenna applications. In this case, a tradeoff between the filter rejection level and its size was necessary. To significantly improve the control of mutual coupling or interference between signals, it is often enough to reach rejection levels above 20–25 dB [2,3,28,29]. Typically, in radiant systems, the isolation or filtering level is reinforced by other effects produced by the antenna itself, such as its impedance as a function of frequency, its position, or its polarization. While on the other hand, a size between half a wavelength and one and a half seems adequate for the objective of integrating the filters in the feeding networks of the antennas. To achieve enough filtering levels in the applications, it was decided to implement a filter with five cells. Considering the size of the EBG elements, this number of resonators offers a good rejection level but at the same time allows the integration of the filter in the feeding network.

### *3.1. Design and Key Parameters: Short-Circuit Element*

The manufacture of the filter requires the implementation of the short-circuit elements of the resonators. In microstrip technology, a common way of making them is through metallized via holes in the printed circuit board of the substrate. In our case, given the gap waveguide configuration, this was implemented using screws. They go through the upper plate of the gap waveguide and make electrical contact only by pressure (without soldering) with the resonator patches in the appropriate position. Since we verified the importance of the radius of the short-circuit element in the working frequency of the resonators, different simple models of the screw were made in the simulation software tool to analyze its effect. Figure 13a includes one of the analyzed models in which the screws are defined as steel cylinders and the nut of the screws is described by a spaced variation of the width of the cylinder. In this case, it can be seen how the thread has a depth given by the difference between the external radius (re) and the internal radius of the screw (ri). Finally, Figure 13b includes the detail of the screw model made for the 5-resonator filter.

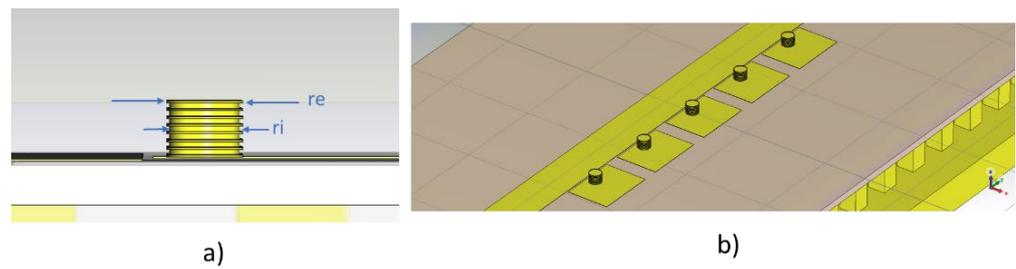

a)    b)

**Figure 13.** Detail of the model implemented for the screw: (**a**) screw thread in the air layer with internal radius (ri) and external radius (re) (**b**) resonators with modeled screws for the 5-cell EBG filter.

For the goals specified in the previous sections, the filter response was adjusted for a frequency of f = 9.5 GHz in the X band. The final structure parameters modeled in the simulator, considering the via as a metallic cylinder, were: rx = 4.8 mm, ry = 6.5 mm, gx = 0.2 mm, gy = 1.75 mm, d = 0.3 mm, and r = 0.6 mm. The S parameters obtained in the simulation for the filter in this case are shown in Figure 14 . In our case, different screws were simulated. Figure 14 shows those corresponding to two screw cases compatible with the M1.4 metric. In this example the external radius of the screw was re = 0.7 mm and two depths for the screw thread were evaluated electromagnetically for ri = 0.6 mm and ri = 0.65 mm. To compare the results with the metallic via models used previously in the work, the results for the same structure but simulated with different radii of the metallic via are also included in Figure 14.

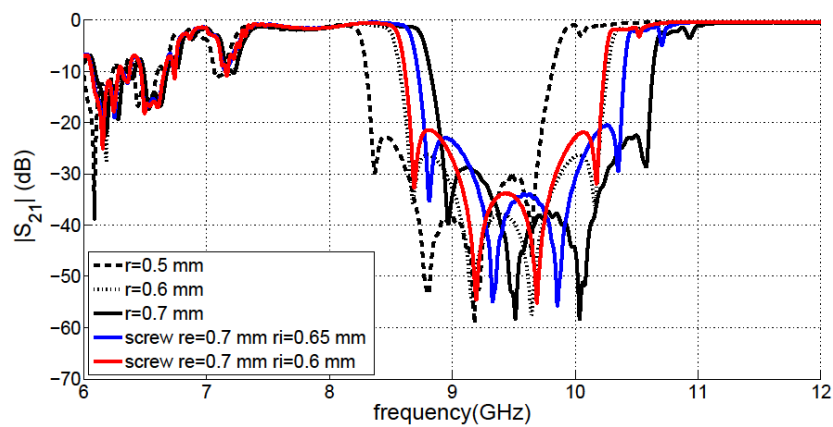

(**a**)

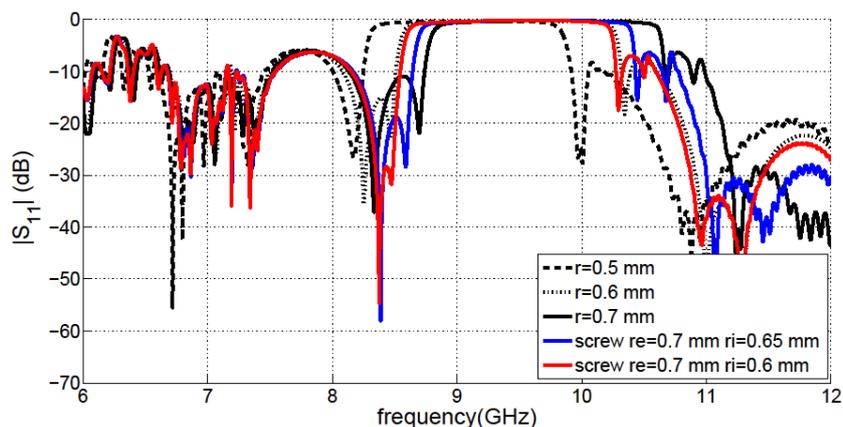

(**b**)

**Figure 14.** **Simulated** S-parameters for the proposed 5-cell EBG-filter for different short-circuit elements (rx = 4.8 mm, ry = 6.5 mm, gx = 0.2 mm, gy = 1.75 mm, d = 0.3 mm); (**a**) S21 (**b**) S11.

From the simulations we obtained valuable information about the importance of the short-circuit elements in the proposed configuration. On one hand, it was verified that when the radius of the via decreases, it shifts the operating frequency of the filter towards lower frequencies. On the other hand, from the screw models we could state that the operating frequency shifts downward as the internal radius of the screw decreases. For the same external radius, when the depth of the thread is greater, the filter works at slightly lower frequencies. It appears that the depth of the thread is related to a smaller effective radius of the shorting element.

The simulated electric field component Ez of the demonstrator filter is included in Figure 15. The figure includes simulations at three frequencies of interest that perfectly describe the performance of the structure. It can be seen in Figure 15a how at f = 6.5 GHz the gap waveguide was outside the working range of the pin surface as we calculated in Figure 2. Clearly in this case the parallel plate did not prevent mode propagation, and the field distribution showed how the energy was not confined to the line. At a frequency of f = 9.5 GHz the filter was in its stopband, so it avoided the propagation of energy as shown in Figure 15b. Finally, it was verified that outside the stopband of the filter and within the working range of the PMC surface, at f = 11.5 GHz in the Figure 15c, the wave propagated confined to the line. In this case, the energy essentially traveled in the air layer, with low losses, in the conventional way that the inverted microstrip version of gap waveguide technology works.

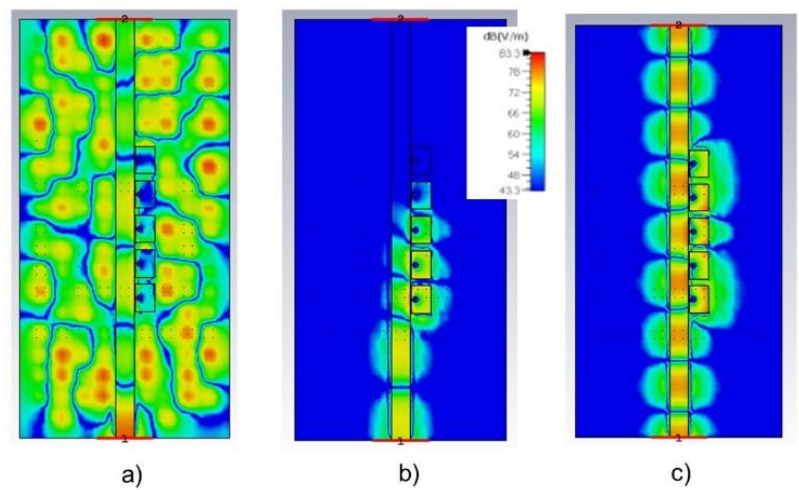

**Figure 15.** Simulated Ez field component for the 5-cell EBG filter prototype at frequencies: (**a**) f = 6.5 GHz (**b**) f = 9.5 GHz and (**c**) f = 11.5 GHz.

*3.2. Manufacture and Measurements*

This section describes the fabricated demonstrator and the measurements made to evaluate the filtering capabilities of the proposed EBG structure. Figure 16 includes some images of the layers necessary for the correct integration of the inverted gap waveguide configuration. It can be seen how the resonator patches were manufactured on the same circuit board as the line, in this case it was necessary to implement a taper on it to properly solder the SMA connectors. The effect of this taper from the electromagnetic point of view was evaluated and was completely negligible. The microstrip line width was w = 4.5 mm to provide the 50 Ω impedance. The demo filter was manufactured with the parameter values proposed at the beginning of the section. The overall dimension of the prototype, including the five resonators, was 39.5 mm × 4.8 mm corresponding to 1.18 λ × 0.14 λ, where λ is the guided wavelength in the air layer of the inverted line at center frequency of 9 GHz. Standardized steel screws with a length of 5 mm and metric M1.4 were used for the integration of the filter, as shown in Figures 16c,d. You can also see in the photographs, included in Figure 16, the holes made in the parallel plate to ensure proper alignment of the layers. Although the gap waveguide technology allowed for easy mechanical integration, in this case it was important to position the layers correctly to ensure that the shorting screws made electrical contact with the resonator patches in the correct position. Finally, it should be noted that in this case the SMA connectors were conventional and were soldered in our laboratory as shown in Figures 16c,d.

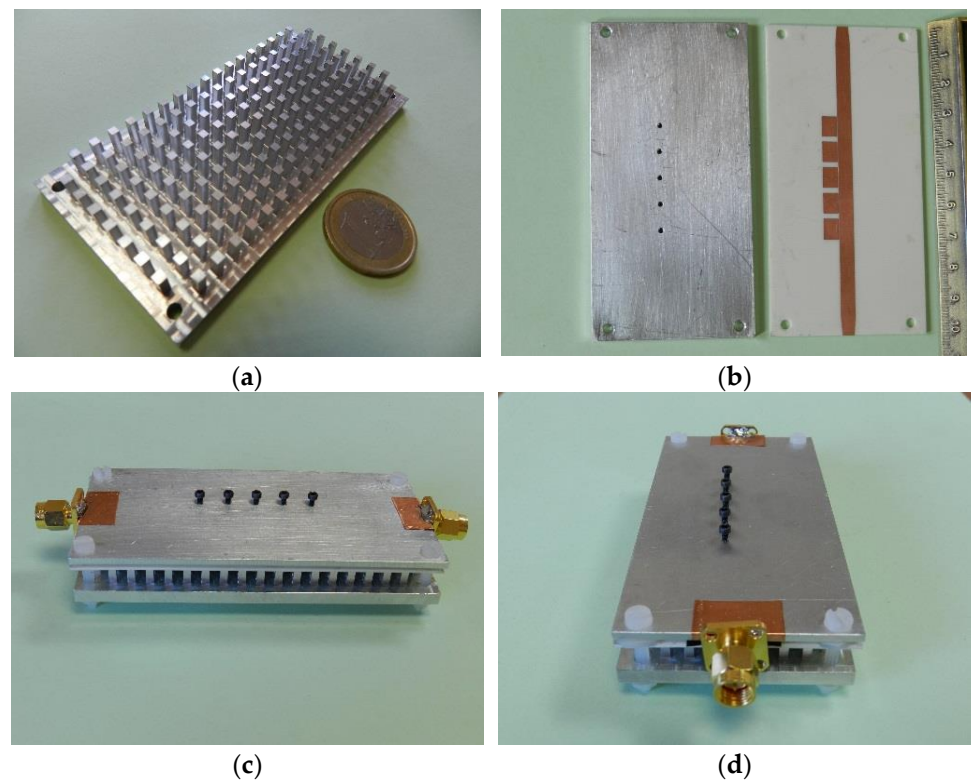

**Figure 16.** Demonstrator manufactured with 5-resonators: (**a**) Bed of nails (**b**) Printed circuit board and top plate in aluminum with the thread for the screws (**c**) General view of the gap waveguide assembly with the short-circuit screws (**d**) Detail of the SMA connectors and soldering.

The manufactured finite EBG structure was characterized experimentally. For this, a vector network analyzer was used in the microwave laboratory. In this type of measurement, it is common to eliminate the effect of the connectors by means of a TRL calibration. In the measurements shown here, this was not carried out because the objective was to evaluate the operation of the EBG filter for its integration in antenna feeding networks. It is evident that the insertion losses obtained in this case were those due to the structure with its materials and to the connectors. Typically, the insertion loss levels of the inverted microstrip gap waveguide technology are between 1 dB and 2.5 dB for frequencies from the Ku to Ka band and can be consulted in the literature [16,17,20,30,38].

Figure 17 shows the measurements of the proposed filter and compares the simulated results for three models, two with short-circuit elements implemented with metal cylinders and another with screws. It can be seen that there was a good agreement between the experimental results and those obtained through simulations with the screw model. Figure 17a shows that the measured central frequency was f = 9 GHz, and the relative fractional bandwidth in the stopband below −20 dB was 10.6%. The measured stopband shifted to lower frequencies, which corresponded to a smaller effective radius of the short-circuit elements than the one used. The measurement for a screw with external radius re

= 0.7 mm (and with ri = 0.65 mm for M1.4 metric screw) corresponded in the simulation to the band obtained for a cylindrical via of r = 0.5 mm. The measured return losses in Figure 17b were 2 dB at a center frequency of 9 GHz while the maximum rejection level of the filter reached -35.4 dB at a frequency of f = 9.03 GHz. In the measurements included in Figure 17, a flat passband was achieved between 10.2 GHz and 12 GHz with return losses above 15 dB for most of this band. The measured insertion losses in this band were between 1.83 dB at f = 10.7 and 1.28 dB at f = 11.2 GHz, including connectors and without TRL calibration. This represents a low level of losses and demonstrates the usefulness of this technology to implement filters in new antenna designs operating in the X and Ku bands.

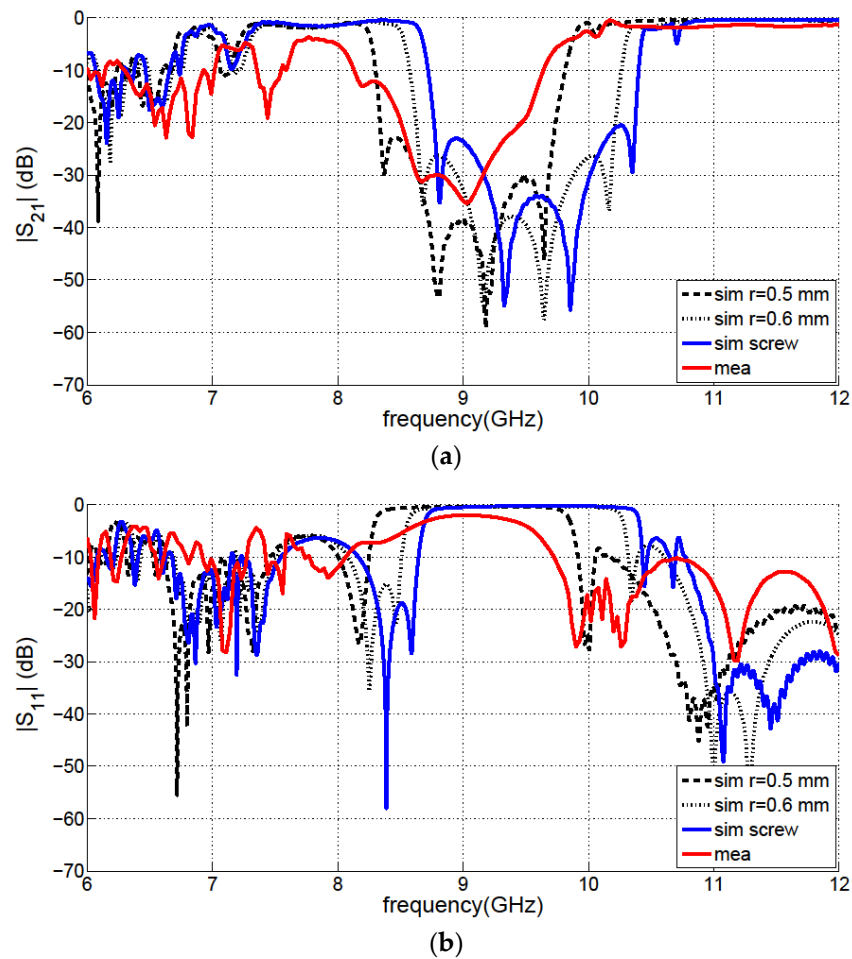

**Figure 17.** Measured (with M1.4 metric screw) and simulated results for the proposed EBG filter (with via of radius r = 0.6 mm and r = 0.5 mm and for the screw model re = 0.7 mm and ri = 0.65 mm) for 5-cell EBG filter: (**a**) S21 (**b**) S11.

*3.3. Discussion*

In this work, the possibility of designing a new inverted microstrip gap waveguide filter has been analyzed from the perspective of easy design and integration. The simulation results showed that by means of an adequate configuration, a compact stopband filter with enough bandwidth can be implemented for X and Ku band antenna systems. It is possible to use a single resonator to generate a frequency notch for narrowband filtering applications such as the suppression of unwanted radiation at specific frequencies to be avoided in ultra-wideband systems. The main design parameters have been analyzed in detail and two of them are of greater importance when considering the design of new devices. The first one is the small coupling gap between the transmission line and the resonator patch. This gap, which at these frequencies must be of the order of a few tens of millimeters, does not present a severe limitation from the manufacturing point of view, but it must be carefully evaluated from the system power handling capability. Similarly, it was verified in the numerical study how this parameter can be useful for increasing the width of the filter stopband or adjusting the working frequency in specific designs. The second one was the radius of the shorting elements that offers interesting alternatives. For example, it increases the flexibility in the designs and enables the reconfiguration of the frequency response of the filters without affecting the topology of the line. This solution can be interesting for filtering applications in planar antenna arrays keeping the initial topology of the feeding network unchanged. The radius of the shorting vias can be used to make the resonators and the filter more compact, making it possible to integrate them with previously designed networks. Finally, the practical implementation of these short-circuit elements produces difficulties related to the greater complexity of the modeling in the simulation when screws are used for their manufacture. In this case, the incorporation of the details of the elements to be integrated in the models and the consideration of their effective radius is important when it comes to obtaining a good agreement between the simulated results and the measurements.

To compare the performance of the proposal, other state-of-the-art filters are included in Table 1. As far as I know, there is no stopband gap waveguide filter implemented in the inverted microstrip version of the technology. For this reason, Table 1 shows a greater variety of filters. On the one hand a comparison is established with some examples of filters based on EBG structures (without using gap waveguide technology) and on the other with some filters that use gap waveguide technology and substrate integrated waveguide technology [39]. In each case, the version of the technology is indicated and if it has stopband or passband filtering functionality.

**Table 1.** Comparison with other filters of the state of the art. Tech: Technology and type, Func: Filter functionality, BW: Fractional bandwidth (rejection above 20 dB for stopband and 3-dB FBW for passband filters), Rejec: Max rejection level for stopband filters, I.Loss/R.Loss: Insertion loss and return loss in the passband, Size: 2D size for filter in guided wavelength.

| Ref. | Tech. | Func. | Freq (GHz) | BW (%) | nº cells/ order | Rejec. (dB) | I. Loss/R. Loss | Size (λ × λ) |
|---|---|---|---|---|---|---|---|---|
| [24] | EBG | Stopband | 10 | 56.78 (−25 dB) | 6 | −42 | Ripple level 1.56 lower and 4.68 upper bands/15 | 0.12 × 2.64 |
| [25] | EBG | Stopband | 7 | 93.7 | 5 | −42 | 0.8 (lower band)/13 | 0.32 × 1.18 |
| [26] | EBG | Stopband /lowpass | 6.5 (cutoff) | - | 6 | −30 | 0.8 (lower band)/20 | 0.39 × 0.91 |
| [39] | SIW | Passband | 9.23–14.05 | 2.5–5.6 | 2 | - | 2.9–2.7 | 1.27 × 2.7 |
| [15] | G-GW | Passband | 11.87 | 3.88 | 6 | - | 0.7/18 | 2.31 (length) |
| [16] | IMR-GW | Passband | 13.57 | 3.76 | 4 | - | 1.41/15.1 | 1.38 × 3.18 |
| [17] | PR-GW | Passband | 31 | 3 | 4 | - | 2.3 | 0.4 × 0.4 |
| [18] | IMR-GW | Passband | 13.17–19.3 | 2–2.4 | 2 | - | 1.6–1.8/10 | 0.61 × 1.48 |
| [20] | IM-GW-SRR | Passband | 28 | 4 | 2 | - | 1.1 | 0.47 × 1.14 |
| [20] | IM-GW-end | Passband | 28 | 11.7 | 4 | - | 1.45 | 1.7 (length) |
| This work | IM-GW-EBG | Stopband | 9 | 10.6 | 5 | −35.4 | 1.5/15 | 0.14 × 1.18 |

The first three papers in Table 1 [24–26] indicate the results obtained by stopband EBG filters on conventional microstrip technology. Their frequencies for the passband are lower than those of this work, so the losses associated with the substrate are still limited. However, these filters present a considerable ripple level that in the upper band reaches values between 5 and 10 dB; this did not occur in our case where the response was very flat. The filters presented in [39] (substrate integrated waveguide SIW in Table 1) and [15] (Groove gap waveguide G-GW in Table 1), although they do not correspond to microstrip technologies, were included to be able to compare the loss levels at frequencies close to 10

GHz. Finally, the rest of the works [16–18,20] included in Table 1 belong to inverted microstrip (IM-GW in Table 1) or to the microstrip-ridge version or similar (IMR-GW and PR-GW in Table 1). All these filters have bandpass functionality, so their specific characteristics are not directly comparable. However, it is interesting to note that both the loss and matching level in the passband are very similar to our proposal. Finally, it is verified how in our case the coplanar EBG filter is compact, which facilitates its integration in small spaces. It is clear from Table 1 that the proposed filter is competitive both in size, rejection level, and losses. This structure can be easily combined in future works with planar gap waveguide antennas to improve their performance.

## 4. Conclusions

This work has presented the analysis of a compact stopband filter implemented by EBG resonators that load an inverted microstrip line in gap waveguide technology. The resonators occupy the same layer as the microstrip and require a metallic shorting element that must be electrically connected to the parallel plate. The effects of the main design parameters on the stopband response of the filter were analyzed. Two parameters were identified, the gap between the patch and the microstrip line and the short-circuit element radius, which are important for adjusting the operating band and allow flexibility in both its size and the characteristic response. A prototype consisting of five EBG cells was manufactured whose short circuits were implemented with screws. The final dimensions for the filter on the printed circuit board were $1.18\ \lambda \times 0.14\ \lambda$. The measured frequency response was in good agreement with simulations including the screw model. The proposed filter had a central operating frequency of 9 GHz and a relative fractional bandwidth in the stopband below −20 dB of 10.6%. The filter reached a maximum transmission rejection level in the stopband of −35.4 dB, and in the passband had a good matching level and low insertion losses of around 1.5 dB. Some of the advantages offered by this inverted gap waveguide technology, low losses, self-packaging, and robustness in mechanical integration, were verified both numerically and with the manufacture of the demonstrator that operated in the X band. The proposed configuration allows filters to be directly integrated, without additional complexity, into transmission lines that are part of the feeding networks of antennas. The results achieved in this work are useful for antenna designs with new filtering capabilities that allow increasing the isolation between ports, reducing interferences, or avoiding unwanted radiation.

**Funding:** This research was funded by the Spanish Ministerio de Ciencia, Innovación y Universidades under project TEC2016-79700-C2-R and by Agencia Estatal de Investigación under project PID2019-107688RB-C21.

**Institutional Review Board Statement: Not applicable**

**Informed Consent Statement: Not applicable**

**Data Availability Statement:** Not applicable.

**Acknowledgments:** I want to thank Santiago, Carola and Quique for their help with the language revision of the document.

**Conflicts of Interest:** The author declares no conflict of interest.